\documentclass[%
 reprint,
 amsmath,amssymb,
 aps,
]{revtex4-1}

\usepackage{graphicx}
\usepackage{dcolumn}
\usepackage{bm}
\usepackage{natbib}
\usepackage{upgreek}
\usepackage{nicefrac}
\usepackage{subfigure}
\usepackage[usenames, dvipsnames]{color}
\usepackage[normalem]{ulem}

\graphicspath{ {./} }

\begin{document}

\title{Effect of ion-trap parameters on energy distributions of ultra-cold
atom-ion mixtures}

\author{Meirav Pinkas}
 
\author{Ziv Meir}
\altaffiliation[Current affiliation: ]{Department of Chemistry, University of Basel, Basel 4056, Switzerland.}

\author{Tomas Sikorsky}
\altaffiliation[Current affiliation: ]{Atominstitut - E141, Technische Universitat Wien, Stadionallee 2, 1020 Vienna, Austria}
\author{Ruti Ben-Shlomi}

\author{Nitzan Akerman}

\author{Roee Ozeri}

\affiliation{%
 Department of Physics of Complex Systems, Weizmann Institute of Science, Rehovot 7610001, Israel
}%

\begin{abstract}
The holy grail of ion-neutral systems is reaching the s-wave scattering regime. However, most of these systems have a fundamental lower collision energy limit which is higher than this s-wave regime. This limit arises from the time-dependant trapping potential of the ion, the Paul trap. In this work, we studied both theoretically and experimentally, the way the Paul trap parameters affect the energy distribution of an ion that is immersed in a bath of ultra-cold atoms. Heating rates and energy distributions of the ion are calculated for various trap parameters by a molecular dynamics (MD) simulation that takes into account the attractive atom-ion potential. The deviation of the energy distribution from a thermal one is discussed. Using the MD simulation, the heating dynamics for different atom-ion combinations is also investigated. In addition, we performed measurements of the heating rates of a ground-state cooled $\ ^{88}$Sr$^+$ ion that is immersed in an ultra-cold cloud of $\ ^{87}$Rb atoms, over a wide range of trap parameters, and compare our results to the MD simulation. Both the simulation and the experiment reveal no significant change in the heating for different parameters of the trap. However, in the experiment a slightly higher global heating is observed, relative to the simulation.
\end{abstract}

\maketitle


\section{\label{sec:intro}Introduction}

Co-trapping ultra-cold atoms and cold ions offers new possibilities
for exploring low-temperature collisions, that include phenomena such
as s-wave scattering, Feshbach resonances \cite{Tomza2015}, shape-resonances \cite{DaSilva2015}
and the creation of molecular ions \cite{Aymar2011}. In addition,
it is also a promising platform for performing quantum computations \cite{Doerk2010}, quantum simulations \cite{Bissbort2013} and for studying out-of-equilibrium dynamics \cite{Meir2018}. In the last decade
these hybrid systems were realized in several experiments, for reviews see \cite{Harter2014,Willitsch2015,Cote2016,Tomza2019}. The interaction
of trapped-ions with ultra-cold thermal clouds \cite{Grier2009} and quantum degenerate gases of neutral atoms \cite{Schmid2010} was studied.
Nonetheless, most of the experiments were limited to atom-ion interaction energy which is greater than the energy scale of the quantum phenomena mentioned above. Only recently, collisions at this energy scale were observed in a system with a heavy ion and light atoms \cite{Feldker2019}.

The interaction energy limitation arises from the fact that the ion is trapped using a Paul trap \cite{Paul1990},
which is based on a time-dependent potential. This potential can inject energy into the system during a collision. Already in 1968, it was observed by Major and Dehmelt \cite{Major1968} that collisions  of ions with heavy atoms in a Paul trap lead to exponential heating of the ion and subsequently to its loss. Forty years later, DeVoe \cite{DeVoe2009} demonstrated numerically that a single collision cannot cause this enormous heating effect,
but it is rather caused by a sequence of collisions which occur at a certain phase of the oscillating trapping potential. He also showed that these consecutive collisions lead to a power-law energy distribution which is not thermal, as one would
expect from a thermalization process. The Tsallis distribution \cite{Tsallis1988}, originally proposed as a generalization of the Boltzmann-Gibbs statistical mechanics to non-extensive systems, was proposed to describe this energy distribution \cite{DeVoe2009}. In addition to a characteristic temperature parameter, this distribution also has a parameter describing its power-law tail. Only recently it was shown, by the formalism of super-statistics, that this distribution indeed arises in the limit of multiple collisions \cite{Rouse2017}.
The power-law tail of the distribution depends on the atom-ion mass ratio as well as the specific Paul trap parameters \cite{Zipkes2011,Goodman2012,Chen2014,Krych2015,Holtkemeier2016,Rouse2018}.

The ion's energy distribution has a characteristic energy scale which can have various sources. If the atoms, colliding with the ion, are at a high temperature, their temperature will determine this energy scale. However, if the colliding atoms are at zero temperature, the energy of the ion can be lost in a single collision if the collision occurs at the center of the trap, ideally leading to a zero energy steady-state for the ion as well. However, it was shown \cite{Zipkes2010,Zipkes2011,Schmid2010} that the mean energy of the ion is typically much higher than that of the atoms by more than an order of magnitude. This additional energy scale arises from the fact that static stray electric fields can move the ion equilibrium position such that it will experience non vanishing rf fields. This effect is called excess micromotion (EMM). Excess micromotion can be compensated by applying an external static electric field that moves the ion to the null point of the rf field \cite{Berkeland1998}.
It was Cetina et. al., in their seminal paper \cite{Cetina2012}, which realized that even without EMM and for an ion in the ground state, a single collision with an atom at zero temperature can still increase significantly the ion's energy. This effect occurs due to the attractive polarization potential between the ion and the atoms that causes the first collision to happen far from the equilibrium position of the ion, and hence at a region with nonvanishing time-dependent electric fields. This heating effect was observed experimentally by Meir et al. \cite{Meir2016a}. There, heating was observed even when the EMM was sufficiently compensated and a cloud of ultra-cold atoms was overlapped with an ion in its ground state. In addition to the heating, a deviation from a thermal distribution was observed.

In this paper we study how the energy distribution and dynamics of an ion, immersed in a bath of ultra-cold atoms, inside a linear Paul trap, and beyond the first collision, depends on different trapping parameters. The heating rates and energy distribution of the ion are calculated by a molecular dynamics (MD) simulation that takes into account the back-action of the polarization of the atom on the ion. This simulation is performed over a wide range of trap parameters as well as for different atom and ion species. In addition, we experimentally measured the heating rates of the ion for different trap parameters and compared our measurements to the results of the MD simulation.

This paper is organized as follows. The MD simulation and its underlying assumptions are described in section \ref{subsec:MD-Model}. The numerical results of the ion distribution dynamics and its dependence on trap
parameters and atom-ion combination are given in section \ref{subsec:num_results}.
In section \ref{subsec:exp_methods} we briefly review the experimental system. The measurements of the heating rates during the first few collisions are presented in section \ref{subsec:exp_results}.

\section{Molecular dynamics simulation}

\subsection{Model\label{subsec:MD-Model}}

Without atoms, the ion's motion depends only on the static and dynamic
confining potentials as described by the Mathieu equation \cite{Leibfried2003},

\begin{equation}
\ddot{u}_{i}+\frac{\Omega^{2}}{4}\left(a_{i}+2q_{i}\cos\left(\Omega t\right)\right)u_{i}=0\label{eq:Mathieu}
\end{equation}

where $u_i$ is the position of the ion in the i-th direction ($i=x,y,x$), $a_{i}$ ($q_{i}$) is the dc (rf) trap parameter and $\Omega$ is the rf drive frequency. In a linear Paul trap, these trap parameters are defined by \cite{Leibfried2003},

\[
a_{x,y}=-\frac{4e}{m_\text{ion}\Omega^{2}}\frac{V_{\text{DC}}}{Z_{0}^{2}}\quad\quad a_{z}=\frac{8e}{m_\text{ion}\Omega^{2}}\frac{V_{\text{DC}}}{Z_{0}^{2}}
\]

\[
q_{x}=-q_{y}=\frac{2e}{m_\text{ion}\Omega^{2}}\frac{V_{\text{RF}}}{R_{0}^{2}}\quad\quad q_{z}=0
\]

where $m_\text{ion}$ is the ion mass, $e$ is electron charge, $V_{\text{DC}}$ and $V_{\text{RF}}$ are the dc and rf voltages on the corresponding electrodes and $R_{0}$ and $Z_{0}$ are constants arising from the geometry
of the dc and rf electrodes, respectively.

In the regime of $\left|a_{i}\right|,q_{i}^{2}\ll1$, the solution to the ion trajectory can be written as,

\begin{equation}
u_{i}\left(t\right)\approx A_{i}\cos\left(\omega_{i}t+\phi_i\right)\left[1+\frac{q_{i}}{2}\cos\left(\Omega t\right)\right],
\label{eq:ion_motion}
\end{equation}

where $\omega_i=\frac{\Omega}{2}\sqrt{a_i+q_i^2/2}$ are the secular frequencies and $A_i$ and $\phi_i$ are the harmonic oscillator amplitude and phase in the i-th direction, respectively.

In the presence of an atom, there is a long-range attractive potential which depends on the relative atom-ion distance, 
\begin{equation}
\label{eq:polarization_potential}
V\left(r\right)=-\frac{C_{4}}{2r^{4}},
\end{equation}with $C_4$ the atom polarizability and $r$ the relative atom-ion distance.
At short distances of few nm, there are electronic exchange interactions which cause a strong repulsion between the particles \cite{Aymar2011}. In principle, the differential cross section depends on the collision energy and the short-range potential. However, in the discussed situation, since the collision energies are relatively high, the differential cross section is angle independent \cite{Zipkes2011}. Hence, in this work, we model the potential as an infinite barrier at a distance of 5nm, and follows Eq. \ref{eq:polarization_potential} otherwise. Since collisions are elastic, when reaching the infinite barrier, the atom and the ion leave in a random direction while conserving the total energy and momentum.

The atom-ion polarization potential is long range. Hence, in order to calculate the ion position, one should integrate the equations of motion of the ion and all the atoms, which is a formidable task when there are many degrees of freedom. However, we can reduce the number of calculations by using the fact that the atomic gas is relatively dilute, and hence the inter-particle distance is large. For example, for a gas with density of $10^{11}\ \frac{1}{\text{cm}^{3}}$ there is approximately one atom in a sphere with a radius of $\sim1\ \upmu$m. For that, we can define an "interaction sphere" with radius $R_{\text{int}}=1.2\ \upmu$m around the equilibrium position of the ion and the position and velocity
of both ion and atom are calculated by solving the equations of motion (EOM),

\begin{align}
\ddot{r}_{\text{ion},i} & =-\frac{1}{m_\text{ion}}\frac{2C_{4}\left(r_{\text{ion},i}-r_{\text{atom},i}\right)}{\left|\boldsymbol{r}_\text{ion}-\boldsymbol{r}_\text{atom}\right|^{5}}\label{eq:Newton_eq-1}\\
 & -\frac{\Omega^{2}}{4}\left(a_{i}+2q_{i}\cos\left(\Omega t\right)\right)r_{\text{ion},i}\nonumber \\
\ddot{r}_{\text{atom},i} & =\frac{1}{m_\text{atom}}\frac{2C_{4}\left(r_{\text{ion},i}-r_{\text{atom},i}\right)}{\left|\boldsymbol{r}_{\text{ion}}-\boldsymbol{r}_\text{atom}\right|^{5}}\nonumber ,
\end{align}
where $m_{\text{atom}}$ is the atom mass.

Multiple collisions are simulated by introducing atoms one after
another into the interaction sphere. For each atom, we solve the equations
of motion (\ref{eq:Newton_eq-1}) using fourth-order Runge-Kutta method
until it exits from the interaction sphere. Then, the position
of the ion is evaluated by Eq. (\ref{eq:ion_motion}) until the next atom enters the interaction sphere.

The atoms enter the interaction sphere with a rate $\Gamma_{\text{atoms}}=n\sigma_{\text{atoms}} v_{\text{th}}$
where $n$ is the atomic density, $\sigma_{\text{atoms}}=\pi R_{\text{int}}^{2}$ (the cross
section of a rigid sphere) and $v_{th}=\sqrt{\frac{8k_{\text{B}}T_\text{a}}{\pi m_{\text{atom}}}}$
is the thermal velocity of atoms at temperature $T_\text{a}$. The interaction sphere radius
must be larger than the amplitude of the ion motion. The initial value
of $R_{\text{int}}$ is taken to be $1.2\ \upmu$m. For a typical secular frequency of $\omega_i/2\pi \sim \ 1$ MHz, this radius corresponds to an ion with 
energy of $\sim 300$ mK. If the ion has a comparable amplitude to $R_{\text{int}}$
after a collision, the interaction sphere radius is increased and
remains at the same size until the end of the calculation for that realization.

Since the atoms density is approximately uniform over the ion's trajectory, we sample the entry point of the atoms uniformly on the sphere. The velocity vector is sampled assuming the atom enters from the south pole, $-\mathbf{\hat{z}}$, and then is rotated to the chosen position. The velocity distribution of the atoms is thermal, but the velocity component in the radial direction must be positive, directed only into the sphere. Therefore, the velocity amplitude distribution is not the regular Maxwell-Boltzamnn distribution. In order to reproduce the correct velocity distribution of the atoms, the amplitude of the velocity, $v$,
and the azimuth angle, $\varphi$, and the polar angle (measured relative to $+\mathbf{\hat{z}}$), $\theta$, are sampled from the following distributions,

\begin{align*}
p_{T}\left(v\right)=\frac{2m_{atom}}{\left(k_{B}T\right)^{2}}v^{3}e^{-\frac{m_{atom}v^{2}}{2k_{B}T}} & \quad v\in\left[0,\infty\right]\\
p\left(\theta\right)=\sin\left(2\theta\right) & \quad\theta\in\left[0,\frac{\pi}{2}\right]\\
p\left(\varphi\right)=\frac{1}{2\pi} & \quad\varphi\in\left[0,2\pi\right].
\end{align*}

The amount of calculations that are needed can be reduced further by taking into account
the characteristic length scale at which the polarization force becomes larger than the ion trap force, $R_{0}=\left(\frac{C_{4}}{\omega^{2}m_{\text{ion}}}\right)^{\nicefrac{1}{6}}$\cite{Cetina2012}.
If the minimal distance between the ion and the atom neglecting the
polarization potential is much larger than $R_{0}$, there is no close
contact and the change to the ion's energy is negligible. Therefore,
a full calculation is performed only for atoms that would approach
the ion sufficiently close without taking the polarization potential into account. In addition, the full calculation of the EOM is stopped once the atom
moved away from the ion by at least $R_{0}$. For the $\ ^{88}$Sr$^{+}-\ ^{87}$Rb system
with typical values of $q\sim0.1$ and $\omega_i/2\pi\sim\ 1$ MHz this characteristic
length-scale is $R_{0}\approx64$ nm.

When performing the full integration of the EOM (\ref{eq:Newton_eq-1}) and the atom-ion distance
is less than a critical distance of 5 nm, elastic hard-sphere collision
is assumed and the atom and ion separate at some random angle. Depending
on the rf phase, a temporary bound state can be created as the atom
may not have enough energy to escape from the polarization potential. In this case, it collides several more times until
it gains enough energy \cite{Cetina2012}.

This process is repeated until the elapsed time from the first collision reaches the total interaction time. In order
to get the ion energy distribution, many realizations with random initial
conditions of the ion are calculated.

\subsection{Results\label{subsec:num_results}}

\subsubsection{Langevin rate for a trapped particle}

\begin{figure}
\includegraphics[width=1\columnwidth]{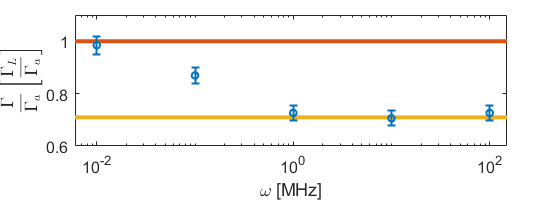}
\caption{\label{fig:Langevin_rate_HO} Langevin rate for atom-ion collisions in a 3D spherical symmetric harmonic trap, normalized to the collision rate of free particles. Each point is an average on $10^4$ repetitions. The ion starts at rest and the atoms have a thermal energy distribution with $T_{\text{a}}=6\ \upmu$K. Red line is the theoretical Langevin rate for two free particles. Yellow line is the limit of an ion with an infinite mass ($\mu = m_\text{a}$).}
\end{figure}

The cross section for an atom which collides with an ion was calculated in 1905 by Paul Langevin \cite{Langevin1905}, 
$$\sigma_\text{L}=\pi\sqrt{\frac{2C_4}{E_{\text{col}}}},$$
where $E_\text{col}$ is the collision energy in the center of mass frame. The collision rate is given by $\Gamma=n\sigma_\text{L}v$ and $v$ is the relative atom-ion velocity.
The MD simulation does not assume the Langevin collision rate \textit{a priori}, and hence it can be calculated. The ratio between the rate of atoms that enter the interaction sphere and the Langevin collision rate is given theoretically by
\begin{equation}
\label{eq:Langevin_ratio}
\frac{\Gamma_\text{L}}{\Gamma_{\text{atoms}}}=\sqrt{\frac{C_4\pi m_\text{atom}}{2\mu k_\text{B} T_\text{a}}}\frac{1}{R_{\text{int}}^2}.
\end{equation}
Numerically, this ratio should be equal to the ratio between the number of hard-sphere collision to the total number of events. In order to check if the collisions rate is changed for a trapped particle, we simulate collisions inside a fictitious 3D time-independent harmonic trap for the ion (without rf fields). As can be seen in Fig. \ref{fig:Langevin_rate_HO}, the resulting collision rate is lower than the theoretical value (red line). In the weak trap limit, the rate almost converges to the ratio in Eq. \ref{eq:Langevin_ratio} since the particle is nearly free. In the strong trap limit, the particle can be regarded as a particle with an infinite mass and then the rate is given again by Eq. \ref{eq:Langevin_ratio} only with $\mu = m_a$, shown by the yellow line.

\begin{figure}
\includegraphics[width=1\columnwidth]{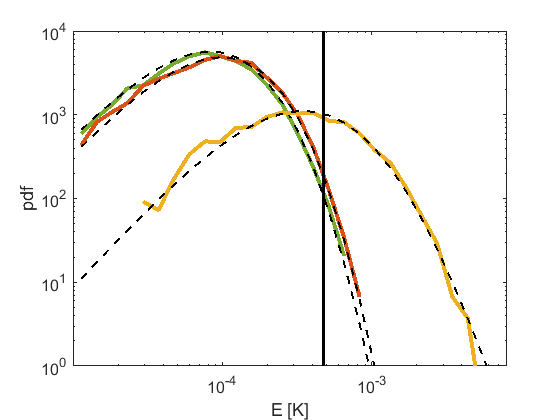}

\caption{\label{fig:first_col_dist} Energy distribution of ${}^{88}$Sr$^+$ ion after a single collision with ${}^{87}$Rb atom as calculated by the MD simulation. The initial energy in each motional mode of the ion is sampled from a thermal distribution of a harmonic oscillator with temperature of $50\ \upmu$K. The temperature of the atomic cloud is $6\ \upmu$K. The red line is the energy distribution before a collision. The yellow line is the energy distribution after exactly one collision in a Paul trap. The green line is the energy distribution after one collision in a fictitious 3D time-independent harmonic trap (without rf fields) with the same secular frequencies as the Paul trap. Dashed lines are fits to a Tsallis distribution. Vertical line is $W_0^{\text{3D}}=472\ \upmu$K (Eq. \ref{eq:W0_Cetina}) taking $q=0.123$ and $\omega$ as the average secular frequency.}
\end{figure}

\subsubsection{Energy distribution after a single collision}
We know that the energy distribution of the ion after many collisions is not a thermal distribution. However, already after one collision, we see that the energy distribution deviates from thermal distribution. In Fig. \ref{fig:first_col_dist}, the energy distributions of the ${}^{88}$Sr$^+$ ion before and after a single collision with a $^{87}$Rb atom in a Paul trap are shown (red and yellow lines respectively). The initial energy in each motional mode of the ion is sampled from a thermal distribution of a harmonic oscillator with temperature of $50\ \upmu$K and the atomic cloud temperature is $6\upmu$ K. The Paul trap frequency is $\Omega/2\pi= 26.5$ MHz and the secular frequencies are $\bar{\omega}/2\pi=(0.821,1.29,0.583)$ MHz in the two radial directions and the axial direction, respectively. The corresponding trap parameters are $\bar{q}=(-0.123,0.123,0)$ and $\bar{a}=(-3.7,1.8,1.9)\cdot 10^{-3}$.

We observe substantial heating after a single collision. A characteristic energy scale for the ion energy gain was derived theoretically in Ref. \cite{Cetina2012}, 
\begin{equation}
\label{eq:W0_Cetina}
W_0^{\text{3D}}=\frac{8}{3\pi}\left(\frac{m_\text{atom}}{m_\text{ion}+m_\text{atom}}\right)^{5/3}\left(\frac{m_\text{ion}^2 \omega^4 C_4}{q^2}\right)^{1/3}.
\end{equation}
For the parameters of the simulation, $W_0^{\text{3D}}=472\ \upmu$K (indicated by the black vertical line), which agrees with the most probable energy of the simulation.

In order to quantify the energy distribution by small number of parameters, we fit to a Tsallis distribution \cite{Tsallis1988},
\begin{equation*}
P_{\text{Tsallis}}\left(E;T,n\right)=\frac{\left(n-3\right)\left(n-2\right)\left(n-1\right)}{2\left(nk_{\text{B}}T\right)^{3}}\frac{E^{2}}{\left(1+\frac{E}{nk_{\text{B}}T}\right)^{n}},    
\end{equation*}
where $E$ is the ion energy, $T$ gives the energy scale (equivalent to temperature) and $n$ describes the power-law tail. This distribution converges in the limit of  $n\to\infty$ to a Maxwell-Boltzmann distribution of a gas in a harmonic trap. In order to compare to a thermal distribution using a single parameter, we define $T_{\text{ion}}=T\frac{n}{n-2}$. This parameter also converges for $n\to \infty$ to the temperature of a thermal distribution.

Fitting the distribution after a single collision to the Tsallis function using Maximum Likelihood Estimation (MLE) gives $T=129\ \upmu$K and $n=7.8$, comparing to $T=44\ \upmu$K and $n=19$ before the collision (equivalent to a thermal distribution). The simulated energy distribution after a single collision in a harmonic trap with the same secular frequencies, is shown by the green line, for comparison. As seen, here the collision does not give a heating effect but cooling, with $T=36\ \upmu$K and $n=16$. This indicates that both the heating and the deviation from a thermal distribution are due to the presence of the rf fields, and occur even after a single collision.

\subsubsection{Energy distribution time dependence}

\begin{figure}
\includegraphics[width=1\columnwidth]{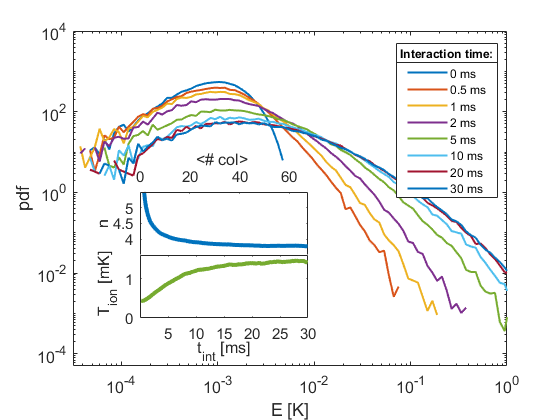}

\caption{\label{fig:distribution_dynamics} $^{88}$Sr$^+$ ion energy distribution after
different interaction times with $^{87}$Rb atoms. The steady-state parameters we find are
$T_{\text{ion}}=1.41\left(1\right)\ $mK, $n=3.77\left(4\right)$. The initial
energy distribution is thermal with temperature of 0.5 mK divided equally
between all motional modes. The Paul trap parameters are identical to the parameters in Fig. \ref{fig:Langevin_rate_HO}. The atomic density $n=1.2\cdot10^{12}$cm$^{-3}$ gives rise to a (numerical)
Langevin rate of $2.25\ \frac{\text{coll}}{\text{ms}}$. Each distribution was constructed
from $5\cdot10^{4}$ repetitions. (inset) Time evolution of the Tsallis
parameters, as calculated by a maximum likelihood fit to the distribution at each interaction time. Confidence bounds are smaller than the width of the lines. }
\end{figure}

To study the heating process dynamics, we sample the energy distribution of the ion at different times. The system is initialized with the same initial conditions as described in the previous section. The time evolution of the ion energy distribution is shown in Fig. \ref{fig:distribution_dynamics}. Not only  the most probable energy is increasing,
but also the high-energies part of the distribution develops a power-law
tail. In addition, as can be also seen in the figure inset, the energy distribution evolution converges to a steady state after $\sim$10 ms, corresponding roughly to 20 collisions.
The values shown in the inset are found by fitting our numerical results to the Tsallis distribution using MLE. Although, as we will show in the following, the Tsallis distribution is only a rough approximation to the ion's energy distribution at steady-state.

\begin{figure}
\includegraphics[width=1\columnwidth]{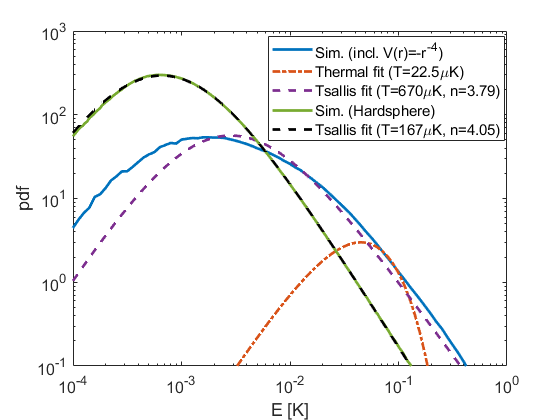}

\caption{\label{fig:MD-steady-state} Ion energy distribution at steady state
for hard-sphere potential (green) and $\sim-\frac{1}{r^{4}}$ potential
(blue) for the same experimental parameters as in Fig. \ref{fig:distribution_dynamics}.
The distribution was calculated for $5\cdot10^{6}$ repetitions. 
Tsallis distribution was fitted using MLE (dashed lines) to all energies
after entering steady-state. In the hard-sphere potential simulation, residual
EMM equivalent to $50\ \upmu$K was added, in the absence of the atom-ion polarization potential.}
\end{figure}

While the Tsallis distribution is an exact limit to the energy distribution of the ion under EMM \cite{Rouse2018}, in the case of an energy distribution generated by a simulation that includes the polarization potential interaction, the Tsallis distribution does not faithfully describe
the distribution. To see this, in Fig. \ref{fig:MD-steady-state}, we compare our steady-state result to 
the steady-state distribution obtained by a simulation with only hard-sphere
potential and EMM, shown by the solid green line, similar to the one detailed in Ref. \cite{Zipkes2011}, which is perfectly described by a Tsallis distribution (dashed black line). However, in the case of a polarization potential simulation (solid blue line) the Tsallis function describes the high energy tail reasonably well, but fails to describe the
most probable energy and the low-energy part (dashed purple line). Nonetheless, it describes the distribution much better than a thermal distribution (dashed red line) and therefore was used in this work to compare energy distributions in different trap parameters.

\subsubsection{Trap parameters dependence}

We now turn to investigate the dependence of the energy distribution on the Paul trap parameters, $a_i$ and $q_i$. These parameters are known to have an effect on the energy gain of the first collision \cite{Cetina2012} and the steady-state power-law \cite{Zipkes2011} in presence of EMM. Here we want to investigate their effect on the steady-state distribution in the absence of EMM. For a linear symmetric Paul trap, the trap parameters are,
\[
a_x=a_y=-\frac{1}{2}a_z=-a \quad
q_x=-q_y=q \quad q_z=0.
\]

\begin{figure}
\includegraphics[width=1\columnwidth]{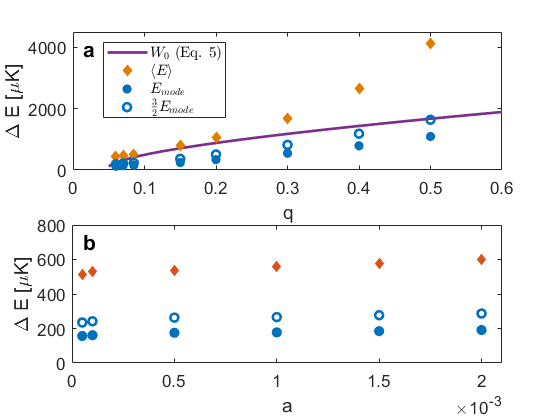}

\caption{\label{fig:Mean-energy-gain}Mean energy gain (red diamonds) and most probable energy ($E_{mode}$, filled blue circles) in the first collision
for (a) different $q$ values and constant $a=-0.001$, and for (b) different $a$ values and constant $q=0.1$. Simulation was performed assuming no EMM. Purple solid line is the characteristic energy scale, $W_{0}$, from \cite{Cetina2012}. Empty blue circles are $\frac{3}{2}E_\text{mode}$, indicating the deviation from a thermal distribution (in which $\langle E \rangle=\frac{3}{2}E_\text{mode}$). }
\end{figure}

The average and mode energy gain in the first collision, which we find in the simulation, are shown in Fig. \ref{fig:Mean-energy-gain}. As seen, the energy gain in the first collision depends strongly on the rf voltage, through the parameter \emph{q}, but shows almost no dependence on the dc confinement, characterized by the parameter \emph{a}. This indicates that stronger rf fields can transfer more energy to the ion during the collision, whereas the amplitude of harmonic pseudo-potential has less dominant impact on the heating. In addition, for a thermal distribution, there is a constant ratio between the mean of the distribution to its mode, $\langle E \rangle = \frac{3}{2}E_\text{mode}$. However, here we can see that this is not satisfied, which is an indication of the non-thermal behavior of the system.

The quantity $W_0$ (Eq. \ref{eq:W0_Cetina}) is indicted in a purple line in Fig. \ref{fig:Mean-energy-gain}. This formula agrees with our simulation for low \emph{q} values, but deviates from our observations for \emph{q} larger than $\sim 0.3$. This might be due to the fact that this energy scale was derived in the absence of dc potentials.

\begin{figure}
\includegraphics[width=1\columnwidth]{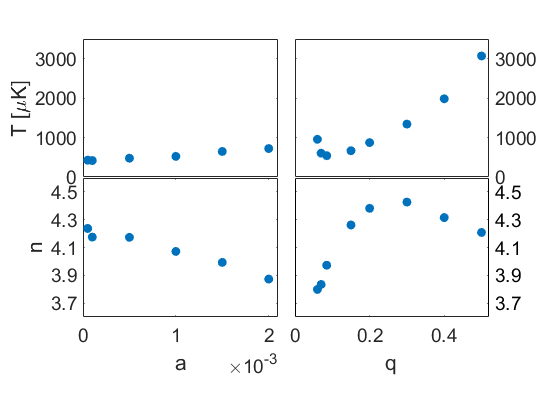}

\caption{\label{fig:Steady-state-a_q}Steady-state Tsallis distribution parameters for Sr$^{+}-$Rb system for different trap parameters of the Paul
trap. (left) different rf confinements with constant dc ($a=0.001$).
(right) different dc confinements with constant rf $\left(q=0.1\right)$.
All values were calculated by preforming MLE to Tsallis distribution.}
\end{figure}

In Fig. \ref{fig:Steady-state-a_q} the dependence of the steady-state distribution as function of the same trap parameters is shown. The distribution is described here by the Tsallis distribution parameters, \emph{T} and \emph{n}, that are extracted from a MLE fit to the simulation results. As before, the rf voltage has a greater influence on the distribution than the dc voltage. Tighter rf confinement leads to higher temperature and lower \emph{n}. Lower \emph{n} means heavier power-law tail and a stronger deviation from a thermal distribution. Similarly to the effect on the first collision, the dependence on the dc parameter is weaker. On the limit of weak rf voltage the distribution is also tending to be hotter with heavier power-law tail. Tיhis can be attributed to the lower spring constant in the radial directions that leads to increased heating.

\begin{figure*}
  \centering
  \subfigure{\includegraphics[scale=0.6]{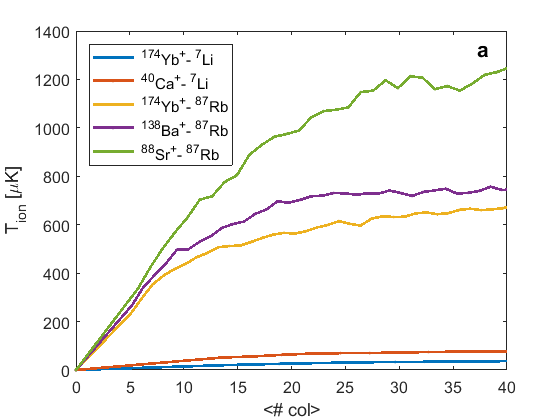}}\quad
  \subfigure{\includegraphics[scale=0.6]{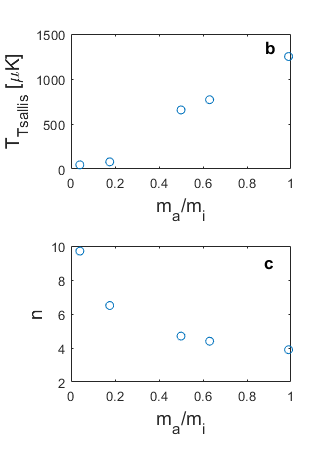}}
  
  \caption{\label{fig:mass-dependence}Time evolution of the most probable energy (a) and steady-state Tsallis parameters: T (b) and n (c) for different atom-ion choices. The parameters $a$, $q$ and $\Omega$ are identical in all realizations. The initial ion temperature is $1\mu K$. The energy distribution was fitted to Tsallis distribution using MLE.}

\end{figure*}
\subsubsection{Atom-ion combinations}

So far we have shown the results of the simulation of our own mixture of $ ^{88}$Sr$^+-^{87}$Rb. Different atom-ion combinations are expected to have different heating rates and steady-state distributions \cite{Chen2014}. It was shown \cite{Cetina2012} that choosing lighter atoms and a heavy ion reduces the effect of pulling the ion from the rf null during the collision. We therefore performed the numerical simulation for several atom-ion systems: $\ ^{174}$Yb$^{+}-\ ^{7}$Li,$\ ^{40}$Ca$^{+}-\ ^{7}$Li,$\ ^{174}$Yb$^{+}-\ ^{87}$Rb,$\ ^{138}$Ba$^{+}-\ ^{87}$Rb and $\ ^{88}$Sr$^{+}-\ ^{87}$Rb.
All other parameters, the dc and rf confinement and trap rf frequency, were
kept constant. In Fig. \ref{fig:mass-dependence}, we can see that
choosing a lower atom-ion mass ratio indeed improves the energy distribution
in two ways: the characteristic temperature, $T$, is lower, and the power-law
exponent, $n$, is larger, leading to lower probabilities for high energy events. The lower temperature in systems with $\ ^{7}$Li atoms is also due to the reduced polarization of the $\ ^{7}$Li relative to the $\ ^{87}$Rb \cite{HandbookChemistry}. The difference in the polarizability and the reduced mass in the center-of mass frame increase the s-wave energy threshold. As can be seen in Table \ref{tab:temerature-steady-state-s-wave}, for systems with $\ ^{87}$Rb atoms the s-wave energy threshold is roughly four or five orders of magnitudes lower than the steady-state temperature whereas systems with $\ ^{7}$Li the threshold is comparable with the obtainable steady-state energies \cite{Furst2018,Feldker2019}.

\begin{table}[]
    \centering
    \begin{tabular}{|c||c|c|c|c|}
    \hline
    Ion-Atom & $\mu\left[\text{amu}\right]$ & $T_{\text{ion}}\left[\upmu \text{K}\right]$ &
    $E_{\text{col}}\left[\upmu \text{K}\right]$ & $E_{\text{s}}\left[\upmu \text{K}\right]$\\
    \hline\hline
    $\ ^{174}$Yb$^+-\ ^7$Li & 6.7 & 38 & 1.5 & 6.4\\
    \hline
    $\ ^{40}$Ca$^+-\ ^7$Li  & 6   & 78 & 11.6 & 8.18 \\
    \hline
    $\ ^{174}$Yb$^+-\ ^{87}$Rb & 58 & 680 & 227 & 0.0443\\
    \hline
    $\ ^{138}$Ba$^+-\ ^{87}$Rb & 53.4 & 740 & 286 & 0.052 \\
    \hline
    $\ ^{88}$Sr$^+-\ ^{87}$Rb & 43.7 & 1200 & 596 & 0.0778\\
    \hline
    \end{tabular}
    \caption{Comparison of steady-state energy for various atom-ion species. For every combination, $T_{\text{ion}}=T\frac{n}{n-2}$ from a fit to the steady-state distribution of the simulation, $E_{\text{col}}$ is the collision energy in the center of mass frame and $E_\text{s}$ is the s-wave energy limit for comparison.}
    \label{tab:temerature-steady-state-s-wave}
\end{table}

\section{Experiment}

\subsection{Methods \label{subsec:exp_methods}}

A full review of our experimental system is given elsewhere \cite{Meir2017}. Briefly, our system consists of two connected vacuum chambers. In the upper
chamber the atoms are collected and cooled to mK temperature
by a magneto-optical-trap (MOT) and then evaporativly cooled in a CO$_2$ quasi-static dipole trap to $\upmu$K temperature. Subsequently, atoms are loaded into a 1D lattice created by two counter-propagating
YAG laser beams. The lattice transfers the atoms to the lower chamber where the ion is trapped. In the lower chamber, the atoms are transferred into a
crossed-dipole trap. The position of the crossed-dipole trap is controlled by a PZT-controlled mirror. After the ion is spin-polarized and ground-state cooled, the crossed dipole trap is moved so that the atomic cloud overlaps the ion. After a given interaction time the atoms are released from the trap and following a short time-of-flight (TOF) are imaged. From the TOF images, the density and the temperature of the atoms are extracted.

The ion is trapped in a segmented linear Paul trap with controlled static
(dc) and dynamic (rf) potentials. The ion is initially Doppler
cooled, and then ground-state cooled using resolved
side-band cooling on the quadruple transition 4d$^2$D$_{5/2}-$5s$^2$S$_{1/2}$.

In order to reduce the heating due to EMM, it is compensated by
applying external electric fields and minimizing the coupling to the EMM resolved
side-bands of the quadruple transition \cite{Meir2017}. This process was performed
before each experiment and periodically every $\sim$40 minutes during
the experiment.

\begin{figure*}

 \centering
\subfigure{\includegraphics[scale=0.5]{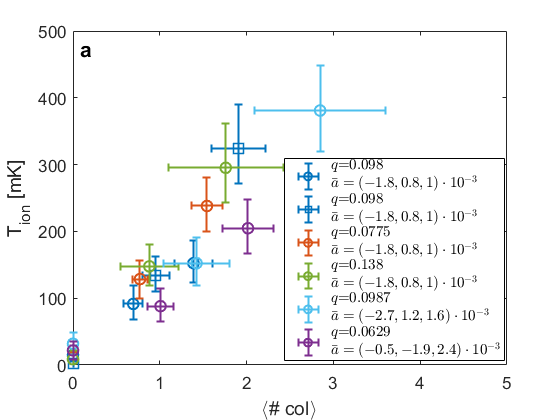}} \quad
\subfigure{\includegraphics[scale=0.5]{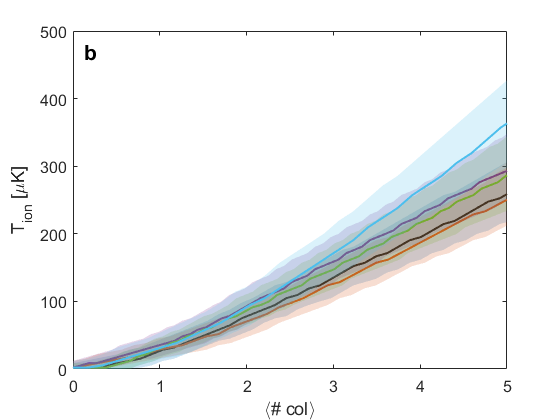}}

\caption{\label{fig:heating-experiment}Measured heating rates for different trap parameters. (a) Temperatures
for different mean number of collisions for different trap parameters as measured in the experiment. Each temperature was extracted from
the Rabi flop data by fitting to a Tsallis distribution with $n=4$.
(b) The heating rates as given by the molecular dynamics simulation. At each time the expected Rabi flop was calculated from the numerical distribution
and then fitted to a Tsallis (n=4) distribution. Error bars of $1\sigma$ are calculated from the likelihood function as described in Appendix \ref{subsec:MLE}.
}

\end{figure*}

\subsection{Results\label{subsec:exp_results}}
Due to experimental limitations, we cannot test the full \emph{a}-\emph{q} space which was simulated. First, we are limited with the maximal values of $a\lesssim2\cdot 10^{-3}$ and $q\lesssim0.15$ due to possible voltage breakdown between adjacent electrodes. On the other hand, in order to perform a ground-state cooling and a carrier thermometry, the ion should be in the Lamb-Dicke regime. This sets lower bounds on the secular frequencies, and hence also on the trap parameters. In our case, a lower limit of $\sim 100$ kHz for the secular frequencies, implies $q\gtrsim0.045$ and $a\gtrsim10^{-4}$. Second, due to a single Doppler-cooling beam used in the experiment, we need to break the radial symmetry in our trap. We do so by applying an additional dc voltage which creates a frequency difference of $\sim100$ kHz between the two radial modes.

The observed temperatures for different interaction times in different trap
parameters are shown in Fig. \ref{fig:heating-experiment}(a).
At each interaction time, a Tsallis temperature was extracted from
the Rabi nutation data by MLE assuming a constant power-law (see Appendix \ref{subsec:carrier_thermo}). As seen, in all experiments the heating rate is roughly the same within the experimental error, 
and on the order of 100-200$\frac{\upmu \text{K}}{\text{ms}}$. The numerical simulation with the exact 
experimental parameters predicts minor differences between different parameters, but all within the confidence bound (see Fig. \ref{fig:heating-experiment}). However, the heating rate in the simulation is much lower and around $50\frac{\upmu \text{K}}{\text{ms}}$.

\begin{figure}

 \centering
\includegraphics[width=1\columnwidth]{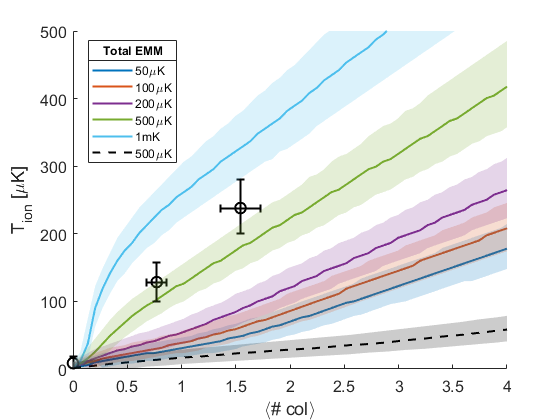}

\caption{\label{fig:heating_sim_EMM}Heating rates from simulations with different residual EMM. Solid lines: simulation that consider polarization potential and various amount of EMM. Dashed line: Simulation with hard-sphere potential and $500\ \upmu$K EMM. Circles refer to the red experiment in Fig. \ref{fig:heating-experiment} (lowest rf experiment).
For the simulation data, error bars are $1\sigma $ from MLE with Tsallis energy distribution where $n=4$. }
\end{figure}

This discrepancy can be explained by several reasons. First, it can arise from a systematic error in estimating the number of atoms through absorption imaging. Another systematic error can arise from the fact that the numerical simulation did not take into account the remaining EMM after compensation. This systematic may vary between different experiments (see Table \ref{table:experiment_EMM} in the supplemental material for estimations). Adding the residual EMM to the simulation changes significantly the heating rates, as can be seen in Fig. \ref{fig:heating_sim_EMM} for one of the experiments. The observed results can be explained by an additional uncompensated EMM of $\sim 500\ \upmu$K. However, this heating cannot be explained by the effect of uncompensated EMM alone, without the polarization potential, which would give much lower temperatures (dashed line in Fig. \ref{fig:heating_sim_EMM}). The effect of heating due to the residual EMM is added to the dominant heating caused by the polarization potential pulling. Although we cannot bound EMM in the system below $\sim 1$mK from our compensation calibrations, our actual EMM is probably at most few 100's $\upmu$K, since otherwise the observed temperatures would be an order of magnitude higher.

\section{Discussion}
Previous studies have shown different heating mechanisms in ultra-cold atom ion collisions. After eliminating those mechanisms, the inherent heating effect of ion pulling from the zero rf point was shown to have a dominant heating effect on the first collision. Here we studied how this process depends on the different choices of Paul trap parameters, and how it affects the ion energy distribution after few to many collisions. We have used a numerical simulation to gain and compare the energy distribution in steady state and for different atom-ion species, a task that can be experimentally hard to perform. We have shown that the heating effect is weakly dependant on the trap parameters, and for experimental purposes is practically independent. We have also shown that the ion's energy distribution clearly deviates from a thermal one. The distribution features a power-law tail, tough it is not described adequately by the previously proposed Tsallis distribution. The experimental measurement suggests that the heating due to residual EMM adds up to the heating due to the polarization potential. In order to considerably decrease this heating effect, apart from working with light atoms and heavy ion, a trapping method without oscillating field for the ion is required, for example, optical trapping of the ion \cite{Lambrecht2017}.
\smallskip

This work was supported by the Crown photonics center, the Israeli Science foundation, the Israeli ministry of Science Technology and Space and the European Research council (Consolidator grant 616919-Ionology)

\newpage

\clearpage

\section*{Appendices}

\subsection{Carrier thermometry\label{subsec:carrier_thermo}}

After the interaction with the atoms, the temperature distribution
is extracted from electron shelving on the quadruple transition. For
a carrier transition on the quadruple transition, the shelving probability
of the excited state (``dark state'') for the n-th level of the
harmonic oscillator is given by \cite{Leibfried2003}

\[
P_{D}\left(t_{R};\boldsymbol{n}\right)=\sin^{2}\left(\Omega_{\boldsymbol{n},\boldsymbol{n}}t_{R}\right),
\]
where $\boldsymbol{n}$ is the harmonic oscillator level (in each
of the 3 modes), $t_{R}$ is the time of the 674nm pulse and the coupling
strength $\Omega_{\boldsymbol{n},\boldsymbol{n}}$ is defined as,
\begin{align}
\Omega_{\boldsymbol{n},\boldsymbol{n}} & \equiv\Omega_{0}\left|\langle n|\exp\left\{ i\sum_{j}\eta_{j}\left(\hat{a}_{j}e^{-i\nu t}+\hat{a}_{j}^{\dagger}e^{i\nu t}\right)\right\} |n\rangle\right|\label{eq:Rabi-freq-n}\\
 & =\Omega_{0}\Pi_{j}e^{-\frac{\eta_{j}^{2}}{2}}L_{n_{j}}\left(\eta_{j}^{2}\right)\nonumber .
\end{align}

Here, $L_{n}\left(x\right)$ is the Laguerre polynomial of order $n$,
$\Omega_{0}$ is the rabi frequency at the ground state and $\eta_{j}=k_{j}x_{j}=\frac{2\pi}{\lambda}\sqrt{\frac{\hbar}{2m\omega_{j}}}\cos\theta$
are the Lamb-Dicke parameters for each mode ($\lambda$ - laser wavelength,
$m$ - ion mass and $\omega_{j}$ the harmonic frequency of the j-th
mode). $\Omega_{0}$, $\theta$ and $\omega_{j}$ are measured independently.

Due to the ion's energy distribution, the population
is divided over many harmonic oscillator levels. Then the probability
to be in the excited state is

\begin{equation}
P_{D}\left(t_{R}\right)=\sum_{\boldsymbol{n}}P\left(\boldsymbol{n}\right)\sin^{2}\left(\Omega_{\boldsymbol{n},\boldsymbol{n}}t_{R}\right)\label{eq:shelv_prob}.
\end{equation}

Where $P\left(\boldsymbol{n}\right)$ is the energy distribution.
However, in our case the probability is given as a function of the
total energy $P\left(E\right)$. In the classical limit when $\bar{n}\gg1$
the energy in the $i$-th mode with $n$ phonons can be approximated
as $E_{i}\left(n_{i}\right)\approx\hbar\omega_{i}n_{i}$. The summation
over the n's is preformed by taking logarithmic spaced n's for each
mode up to some cutoff value, then the corresponding Rabi frequency
is calculated by Eq. (\ref{eq:Rabi-freq-n}). The probability for
this term is taken from the predefined $P\left(E\right)$ probability
(Thermal, Tsallis or numerical). Then, for any given pulse time $t_{R}$
the shelving probability is expressed as an integral over the energy

\begin{equation} 
P_{D}\left(t_{R};P\left(E\right)\right)=\int_{E}P\left(E\right)\sin^{2}\left(\Omega\left(E\right)t_{R}\right) dE.
\label{eq:shelv_prob_energy}
\end{equation}

Qualitatively, a ``cold'' ion will give a high contrast sine-square
function, since only few spectral components are dominant, whereas
``hot'' ion will give dephasing sine-square with faster decay as the temperature is higher.

\bigskip

\subsection{Maximum likelihood estimation\label{subsec:MLE}}
For the experiment, the temperature at a specific interaction time is extracted from the data using Maximum likelihood estimation (MLE).
For each interaction time with the atoms, a Rabi flop is taken at
five different pulse time $t_{i}$. For each time, the experiment is
repeated $N_{i}$ times and $x_{i}$ dark events were observed. The number of dark events has a binomial distribution with probability $P_{D}\left(t_{i};P\left(E\right)\right)$ and $N_{i}$ number of trails.
Therefore, the likelihood of a specific experiment result is
\begin{widetext}
    \begin{equation}
    L\left(P\left(E\right)|\left\{ x_{i},N_{i}\right\} \right)=\prod_{i}\binom{N_{i}}{x_{i}}P_{D}\left(t_{i};P\left(E\right)\right)^{x_{i}}\left(1-P_{D}\left(t_{i};P\left(E\right)\right)\right)^{N_{i}-x_{i}},
    \label{eq:likelihood}
    \end{equation}
\end{widetext}
In order to simplify the calculations, for the experimental data,
the energy distribution for calculating the Rabi flop in Eq. \ref{eq:shelv_prob} is assumed to be a Tsallis distribution with a constant power-law
$n=4$. Hence, the fitting problem is reduced to finding of a single parameter, the temperature $T$, which maximizes the log-likelihood function. Since the likelihood function has a Gaussian shape, confidence bounds, $T_\pm$, of $1\sigma$ are found from the condition \cite{Bohm2010}
\begin{equation*}
    \log L(T_\pm)-\log L(T_\text{max})=-\frac{1}{2},
\end{equation*}
where $T_\text{max}$ is the maximum of $\log L$.
For the MD simulation, at each interaction time, a Rabi flop was calculated by Eq. \ref{eq:shelv_prob_energy}, with the numerical energy distribution given by the simulation. The MLE and the confidence bounds are calculated in the same method as before.

\subsection{EMM estimation}

The process of detecting and compensating excess micro-motion (EMM) is described in detail in Ref. \cite{Meir2017} and will discussed here only briefly. The EMM arises as a result of non-vanishing rf field in the minimum of the pseudo-potential, for example, due to a uniform dc field. For a static electric field $E_{dc}$, the ion motion has an additional term oscillating in the rf frequency. Adding this term to Eq. \ref{eq:ion_motion} \cite{Berkeland1998},

\begin{equation}
u_{i}\left(t\right)\approx\left[\frac{eE_{dc,i}}{m\omega_i^2} + A_{i}\cos\left(\omega_{i}t+\phi\right)\right]\left[1+\frac{q_{i}}{2}\cos\left(\Omega t\right)\right].
\end{equation}
The amplitude of the EMM,
\begin{equation}
    u_{EMM,i}=\frac{q_ieE_{dc,i}}{2m\omega_i^2}.
\end{equation}
In the resolved sideband spectroscopy, this motion causes additional sidebands in the rf frequency. The relative coupling between the carrier and the first EMM sideband is given by,

\begin{equation}
    \frac{\Omega_1}{\Omega_0}\approx\frac{\boldsymbol{k}\cdot\boldsymbol{u}_{EMM}}{2},
\end{equation}

where $\Omega_0$ ($\Omega_1$) is the carrier (sideband) Rabi frequency and $\boldsymbol{k}$ is the laser wave-vector.
\begin{table}
    \begin{center}
     \begin{tabular}{|c | c |} 

     \hline
     Experiment name & $T_{EMM} \big[\mu K\big] $ \\ [0.5ex] 
     \hline\hline
     RF 18 (\#1) & 860$\pm$250  \\ 
     \hline
     RF 18 (\#2) & 700$\pm$80 \\
     \hline
     RF 16 & 1300$\pm$270  \\
     \hline
     RF 21 & 330$\pm$50 \\
     \hline
     DC 390V & 850$\pm$160\\
     \hline
     DC 600V & 1600$\pm$230  \\ [1ex] 
     \hline
    \end{tabular}
    \caption{\label{table:experiment_EMM}EMM energy estimation in the radial directions for each experiment. The energy was estimated by comparing the shelving probability on the EMM-sideband transition relatively to the carrier transition. The EMM that enters from the stability of the electrodes is negligible in all experiments.}
    
    \end{center}
\end{table}
Minimization of the EMM is done by applying an external electric field and minimizing the Rabi frequency. The EMM is mainly in the radial plane of the trap, and therefore the minimization is done with two orthogonal laser beams and two orthogonal electrodes giving constant electric field in the trap center. The EMM in the axial direction due to static electric field is negligible by trap design. However, rf fields can still persist in the axial direction and they are compensated using a rf fields injected along the axial direction with a controllable phase, relative to the rf of the trap. The EMM in terms of energy is given by,
\begin{equation}
    k_B T_\text{EMM}=\frac{1}{4}m\Omega^2u_\text{EMM}^2
    =\frac{1}{4}m\Omega^2\left(\frac{2\Omega_1}{k\cos\theta_i\Omega_0}\right)^2,
\end{equation}
where $\theta_i$ is the angle between the laser beam to the radial plane (measured independently).

Our residual EMM energies after compensation for each experiment are summarized in Table \ref{table:experiment_EMM}. The confidence bounds are derived from the projection noise of the shelving probability after compensation. The extracted energies are considerably large than expected. This can be attributed to different sources which are not EMM. First, the pulse time is relatively long, in order to detect weak coupling to the side band. But, in this timescale the decoherence (for example, due to magnetic field noise), can be dominant. For very low shelving probability, decoherence will increase the population in the excited state, and therefore it will look as EMM.

\end{document}